# The emergent properties of a dolphin social network[1]

David Lusseau

University of Aberdeen, Lighthouse Field Station, George St., Cromarty, IV11 8YJ, Scotland

d.lusseau@abdn.ac.uk

Many complex networks, including human societies, the Internet, the World Wide Web and power grids, have surprising properties that allow vertices (individuals, nodes, Web pages, etc.) to be in close contact and information to be transferred quickly between them. Nothing is known of the emerging properties of animal societies, but it would be expected that similar trends would emerge from the topology of animal social networks. Despite its small size (64 individuals), the Doubtful Sound community of bottlenose dolphins has the same characteristics. The connectivity of individuals follows a complex distribution that has a scale-free power-law distribution for large k. In addition, the ability for two individuals to be in contact is unaffected by the random removal of individuals. The removal of individuals with many links to others does affect the length of the 'information' path between two individuals, but, unlike other scale-free networks, it does not fragment the cohesion of the social network. These self-organizing phenomena allow the network to remain united, even in the case of catastrophic death events.

## 1. INTRODUCTION

Complex networks that contain many members such as human societies (Newman et al. 2002), the World Wide Web (WWW) (Lawrence & Giles 1999), or electric power grids (Watts & Strogatz 1998), have emergent properties that permit all components (or vertices) in the network to be linked by a short chain of intermediate vertices. Recent theoretical and empirical work on complex networks shows that they can be classified in two major categories depending on the likelihood, p(k), that a vertex is linked with k vertices (Albert et al. 2000). The first type of network—the random model described by Erdös and Rényi (1959) and the small world effect of Watts and Strogatz (1998)—has a fairly homogeneous topology, with p(k) following a Poisson distribution that peaks at an average $\langle k \rangle$. The other category, described by Barabási and Albert (Albert et al. 2000), is topologically heterogeneous with p(k) following a scale-free power law for large k that is $p(k) \propto k^{-g}$ (Barabási & Albert 1999). In the first type of model it is unlikely that a vertex has many links and the cohesion of the network is maintained by random 'weak links', in other words links between two individuals belonging to different clusters within the network (e.g., human societies) (Newman et al. 2002). In scale-free networks there exist vertices that act as hubs of activities because they possess many links with other vertices (e.g. the Internet; Barabási & Albert 1999).

Gregarious, long-lived animals, such as gorillas (*Gorilla gorilla*), deer (*Cervus elaphus*), elephants (*Loxodonta africanus*) and bottlenose dolphins (*Tursiops truncatus*) rely on information transfer to utilise their habitat (Janik 2000; Conradt & Roper 2003). Despite some effort to understand how this information is communicated, we still have little understanding of the way these societies are organised to transfer information. I investigated the properties of the social network of bottlenose dolphins (*Tursiops*

---

spp.) present in Doubtful Sound (45°30' S, 167° E), Fiordland, New Zealand.

## 2. METHODS

The Doubtful Sound bottlenose dolphin population is small, 60-65 individuals, and reside year-round in this fjord (Williams et al. 1993). I defined social acquaintances in the network as preferred companionships (Connor et al. 2001), that is individuals that were seen together more often than expected by chance. Every time a school of dolphins was encountered in the fjord between 1995 and 2001, each adult member of the school was photographed and identified from natural markings on the dorsal fin. This information was utilised to determine how often two individuals were seen together. To measure how closely two individuals were associated in the population (i.e. how often they were seen together) I calculated a half-weight index of association for each pair of individuals (HWI) (Cairns & Schwäger 1987). This index estimates the likelihood that two individuals were seen together compared to the likelihood to see any of the two individuals when encountering a school:

$$HWI = \frac{X}{X + 0.5(Y_a + Y_b)}$$

where:

X: number of schools where dolphin a and dolphin b were seen together

$Y_a$: number of schools where dolphin a was sighted but not dolphin b

$Y_b$: number of schools where dolphin b was sighted but not dolphin a

Only individuals that survived the first 12 months of study were considered in this analysis, so that enough information was available to analyse their preferences in association. I tested the significance of these association indices by randomly permuting individuals within groups (20,000 times), keeping the group size and the number of times each individual was seen constant, using SOCPROG 1.3 (developed by Hal Whitehead for MATLAB, available at http://www.dal.ca/~hwhitehe/social.htm). After each permutation the HWI for each pair was calculated and the observed HWI was compared to the 20,000 expected HWI. The number of permutations was not arbitrarily chosen, I increased the number of permutations performed until the p-value obtained from the Monte-Carlo simulation stabilised (Bejder et al. 1998). If more than 95% of the expected HWI were smaller than the observed HWI, the pair of dolphins was defined as a preferred companionship. In other words, the pair of dolphins was more likely to be seen together than by chance.

I compared this social network to random networks that would contain the same number of links and vertices. I investigated the diameter and the clustering coefficient of these networks using UCINET 6 (Borgatti et al. 2002). The diameter, *d*, of a network is defined as the average length of the shortest paths between any two vertices. The smaller *d* is, the quicker information can be transferred between any given individuals. For example, the global human population seems to have a diameter of six meaning that any two humans can be linked using five intermediate acquaintances (Milgram 1967). The clustering coefficient, *C*, gives a measure of the social relatedness of individuals within the network. For each vertex, *n*, it provides the likelihood that two associates of *n* are associates themselves.

## 3. RESULTS

Over the 7 years of observation the composition of 1292 schools was gathered. There were 64 adult individuals in this social network linked by 159 preferred companionships (edges) and therefore the average connectivity of the network, $\langle k \rangle$, was 4.97. The number of links each individual had was not Poisson distributed (goodness-of-fit test: $G^2_{adj, df=12}$ = 26.48, p = 0.009). The tail of the distribution of p(k) was similar to the one of scale-free networks while it seemed to flatten for k<7. The tail of the distribution, $k \geq 7$, seemed to follow a power-law with $\gamma_{dolphin} = 3.45 \pm 0.1$ (Figure 1).

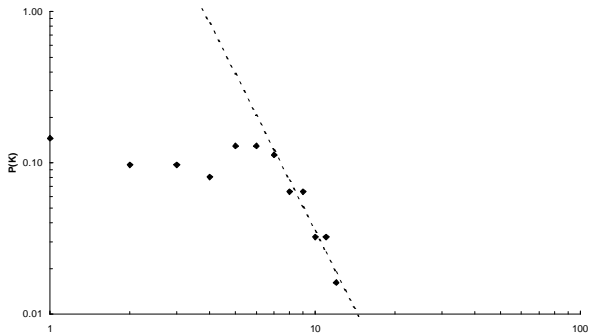

Figure 1. The distribution function of the number of preferred companions (edges, K) for each of the 64 individuals in the dolphin network. There are 159 edges between these dolphins and the average connectivity $\langle k \rangle$ = 4.97. The dashed line has slope $g_{dolphin} = 3.45$.

Both the random networks and the dolphin network had similar diameter (Figure 2, $d_{dolphin}$ = 3.36; $d_{random}$ = 2.72, s.e.$_{(random)}$ = 0.03 over 10 random networks tested), but the dolphin network had a much higher level of clustering ($C_{dolphin}$ = 0.303; $C_{random}$ = 0.081, s.e.$_{(random)}$ = 0.003).

Not surprisingly, the dolphin scale-free network was resilient to random attacks. The diameter of the network increased by only 0.4 with the removal of 20% of individuals (Figure 3). These values are averages of ten different trials to randomly remove vertices. The average mortality rate per year observed from 1995 to 1999 ranged from 1.8% to 7.9% (Haase 2000) so the values tested here were unrealistically high. Targeted attacks on the other hand, that is the removal of individuals with the most associates, affected the diameter of the network (Figure 3). The shortest path between any two given individuals was increased by 1.6 with the removal of 20% of individuals (Figure 3). The dolphin network did not fragment under targeted attacks, but maintained a large cluster encompassing most individuals (Figure 4). Even when more than 30% of individuals were removed, randomly or selectively, the network was characterised by the presence of a large cluster that encompasses most of the individuals present and single individuals without any associates (Figure 4).

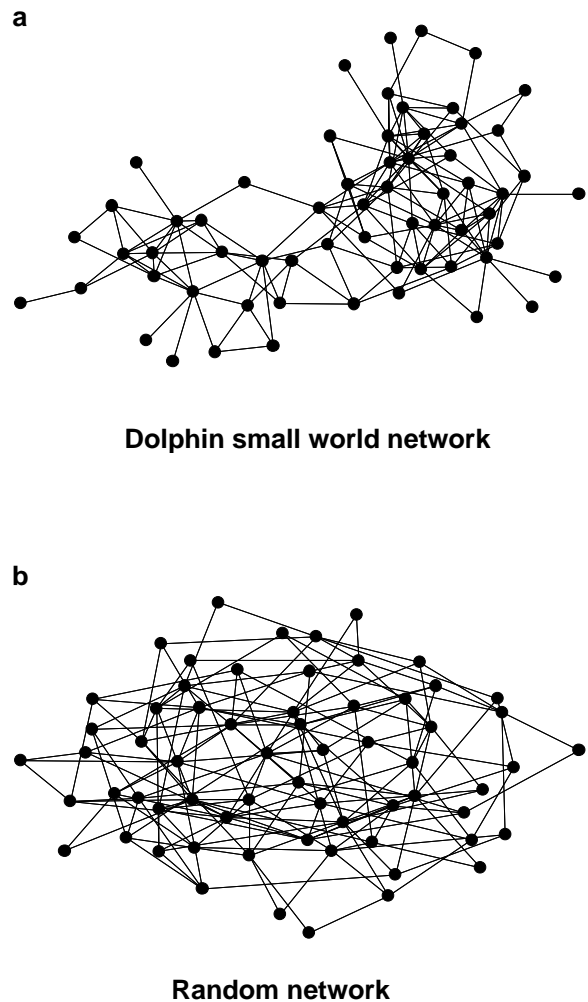

Figure 2. Illustration of the dolphin network and a random network constructed with similar characteristics (N = 64, $\langle k \rangle$ = 4.97). These networks were constructed using Netdraw as part of the UCINET software (Borgatti et al. 2002). **a**, the dolphin network is inhomogeneous, a few vertices have large number of links and many have only one or two links. **b**, the random network is homogeneous, the number of link each vertex has follows a Poisson distribution.

## 4. DISCUSSION

This social network was characterised by the presence of "centres" of associations, which shows that not all individuals have an equal role in this society. These hubs were mainly adult females, at the exception of one adult male, and seemed to be older individuals (many scars and larger size). These individuals seem to play a role in maintaining a short information path between individuals of the population.

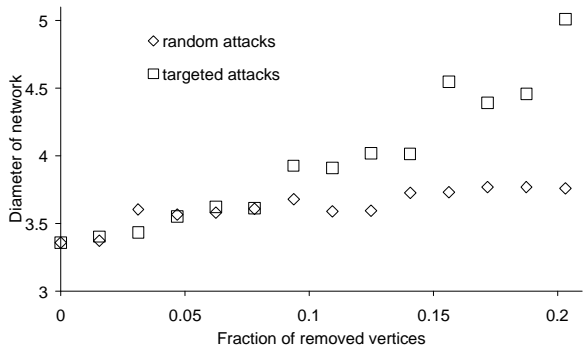

Figure 3. Changes in the diameter of the dolphin social network with the fraction of removed vertices. When the selection of vertices to be removed is random (random attacks) changes in the diameter are minor even after the removal of an unrealistic number of vertices (0.20). When vertices with many links are removed (targeted attacks), the change in diameter is noticeable but differs in magnitude from the behaviour of other small world network under similar attacks (Albert et al. 2000).

The removal of hubs of associations (i.e. individuals with many associates) altered the diameter of the network. However, this increase was trivial compared with previously studied scale-free networks (Albert et al. 2000). For example, the diameter of two large networks (the Internet and the WWW) more than double when 2% of the nodes with the most links were removed (Albert et al. 2000). Random and scale-free networks typically become fragmented into small clusters under targeted attacks (Albert et al. 2000). This was not the case for this dolphin social network. Individuals with many companions do not maintain the cohesion of the network, yet not all individuals in the network play a similar role in its cohesion (Figure 4). The high clustering coefficient of the network may reveal a high level of redundancy in connections. This redundancy would permit to increase the resilience of the network to deaths, by making sure that several short paths exist between any two given individuals in the network. Despite this apparent redundancy in connectivity, $\langle k \rangle$ was well within the range of other scale-free networks ($\langle k_{Internet} \rangle = 3.4$, $\langle k_{WWW} \rangle = 5.46$, and $\langle k_{actors} \rangle = 28.78$; Albert et al. 2000; Barabási & Albert 1999). However, the distribution of link numbers differed as well from typical scale-free networks as it flattened for k<7. Some human social networks have been described with similar distribution characteristics (Barabási & Albert 1999). The resilience of the dolphin network to the removal of individuals may be related to this flattened portion of the distribution. There was no cliques (groups of vertices in which all vertices are connected with each other) with more than five individuals in the network and only three cliques containing five individuals each. It therefore seems that individuals with intermediate number of associates (4-7) play an important role as redundant links between different sections of the network.

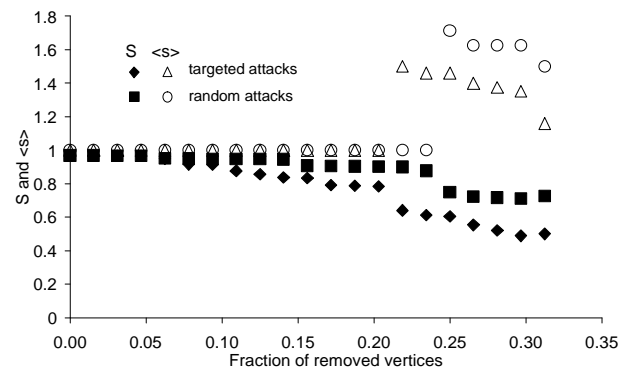

Figure 4. Fragmentation of the network under random and targeted attacks. The size of the largest cluster in the network (S) is relative to the total number of individuals in the network and therefore varies from 0 to 1. The average size of isolated clusters $\langle s \rangle$, clusters other than the largest one, is 1 if all isolated clusters are composed of single isolated individuals, and >1 if isolated clusters are a combination of small clusters containing ≥1 individuals. The fragmentation pattern is similar both under random and targeted attacks. Even after the removal of an unrealistic number of individuals (Haase 2000), the largest cluster contains most of the individuals present in the network (contrarily to other small world networks under targeted attacks).

The benefits behind this emergent resilience are obvious. The resilience properties of this network permit to maintain a cohesive society even in the event of a catastrophe that would result in the loss of more than a third of the population. In addition, the scaling properties are advantageous for a network that evolves with time. They permit to assimilate new vertices without disrupting the cohesion of the network (Barabási & Albert 1999).

This is one of the smallest networks, of any type, in which scale-free emerging properties have been observed. It provides further evidence that these self-organising phenomena do not depend solely on the characteristics of individual systems, but are general laws of evolving networks. The resilience of this dolphin social network to selective and random attacks should be explored further. Such properties could be advantageous to apply to other networks (WWW, the Internet) that can be seriously damaged by targeted attacks (Albert et al. 2000).

**Acknowledgements**

I was the recipient of a University of Otago Bridging Grant during this study. This research was funded by the New Zealand Whale and Dolphin Trust, the New Zealand Department of Conservation, and Real Journeys Inc. The University of Otago (Division of Sciences, Departments of Zoology and Marine Sciences) provided additional financial and technical support. Information on group composition was collected by Karsten Schneider (1995-1997), Patti Haase (1999), Oliver J. Boisseau (2001) and the author (1999-2001). I would like to thank two anonymous referees for their contributions that greatly improved the quality of this manuscript.